\documentclass{article}
\usepackage{spconf,amsmath,graphicx,hyperref}
\usepackage{xcolor}
\usepackage{booktabs}
\usepackage{multirow}

\makeatletter
\def\@name{ 
\emph{Alexander Polok$^{\star}$ 
\qquad Dominik Klement$^{\star}$ 
\qquad Samuele Cornell$^{\ddagger}$ 
\qquad Matthew Wiesner$^{\dagger}$}  \\ 
\emph{Jan Černocký$^{\star}$  
\qquad Sanjeev Khudanpur$^{\dagger}$ 
\qquad Lukáš Burget$^{\star}$}}
\makeatother

\title{SE-DiCoW: Self-Enrolled Diarization-Conditioned Whisper}

\address{$^{\star}$ Speech@FIT, Brno University of Technology, Czechia \\
            $^{\ddagger}$ Language Technologies Institute, Carnegie Mellon University, USA \\
         $^{\dagger}$ CLSP \& HLTCOE, Johns Hopkins University, USA}

\ninept
\begin{document}
%
\maketitle

\begin{abstract}
Speaker-attributed automatic speech recognition (ASR) in multi-speaker environments remains a major challenge. While some approaches achieve strong performance when fine-tuned on specific domains, few systems generalize well across out-of-domain datasets. Our prior work, Diarization-Conditioned Whisper (DiCoW), leverages speaker diarization outputs as conditioning information and, with minimal fine-tuning, demonstrated strong multilingual and multi-domain performance.  
In this paper, we address a key limitation of DiCoW: ambiguity in Silence–Target–Non-target–Overlap (STNO) masks, where two or more fully overlapping speakers may have nearly identical conditioning despite differing transcriptions. We introduce SE-DiCoW (Self-Enrolled Diarization-Conditioned Whisper), which uses diarization output to locate an enrollment segment anywhere in the conversation where the target speaker is most active. This enrollment segment is used as fixed conditioning via cross-attention at each encoder layer. We further refine DiCoW with improved data segmentation, model initialization, and augmentation. Together, these advances yield substantial gains: SE-DiCoW reduces macro-averaged tcpWER by 52.4\% relative to the original DiCoW on the EMMA MT-ASR benchmark.
\end{abstract}
\section{Introduction}
\label{sec:intro}

\begin{figure}[t]
    \vspace{-0.5cm}
    \makebox[\columnwidth][c]{%
        \includegraphics[width=1\columnwidth]{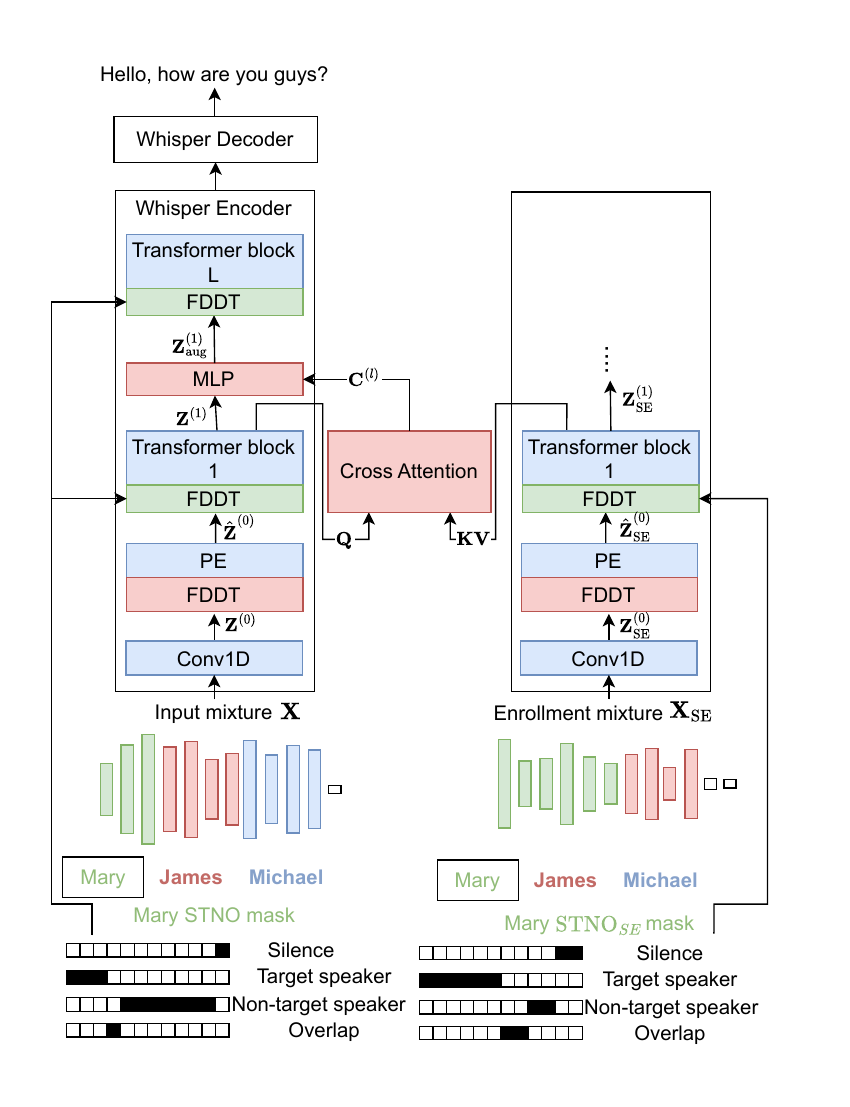}%
    }
    \vspace{-1cm}
    \caption{Overview of the SE-DiCoW model architecture. Newly introduced parameter blocks are highlighted in red.}
    \label{fig:dicow}
\end{figure}

Speaker-attributed automatic speech recognition (ASR) is critical for applications such as meetings, interviews, and other multi-party conversations, where transcripts must capture \emph{who spoke what}. Despite recent advances~\cite{radford2023robust, peng2023owsm, puvvada24_interspeech}, current single-speaker ASR models perform poorly in multi-talker scenarios, struggling with overlapping speech, spontaneous dialogue, and, importantly, failing to provide speaker attribution.
Over recent years, several challenges~\cite{cornell2023chime, cornell2024chime, vinnikov24_interspeech} have driven the development of novel solutions for these difficult conditions. Modular multi-talker ASR approaches that combine diarization, source separation, and ASR~\cite{niu24_chime} have dominated the field, but they are complex, often fail to generalize across domains, and are prone to cascading errors. In contrast, simpler end-to-end strategies, such as speaker-token conditioning~\cite{kanda20b_interspeech, cornell2024one} or multi-decoder architectures~\cite{yu17b_interspeech}, alleviate some of these issues but generally underperform modular approaches.

Target-speaker ASR (TS-ASR) offers a middle ground by directly conditioning ASR models on speaker identity using embeddings or enrollment audio~\cite{Kanda2019_spkloss, ma2024extending}. While effective in controlled settings, these methods often depend on speaker-specific representations~\cite{Zili23_adapting} that are difficult to generalize, particularly when training data is limited, or speaker variability is high.

To address limitations of TS-ASR approaches, we introduced Diarization-Conditioned Whisper (DiCoW)~\cite{polok24_butjhu, polok2024targetspeakerasrwhisper, DiCoW}, a target-speaker ASR framework that conditions Whisper~\cite{radford2023robust} on frame-level diarization masks instead of speaker embeddings. By avoiding explicit speaker-identity modeling, DiCoW scales effectively to real-world conversations with unknown speakers and demonstrates good cross-domain performance. Notably, it outperformed several speech-augmented large language models~\cite{SALM} in a recent multilingual challenge~\cite{polok2025mlcslmchallenge}. The diarization-conditioning paradigm has since been extended beyond Whisper, including its adaptation to the Parakeet-TDT model~\cite{wang25y_interspeech}, as well as DiCoW extensions that explore end-to-end multi-talker modeling with serialized output training~\cite{kocour2025adaptingdiarizationconditionedwhisperendtoend} and inference-time scaling via speaker-agnostic activity streams~\cite{he2025scalingmultitalkerasrspeakeragnostic}.

Despite these successes, DiCoW has a key limitation: while the diarization output is converted into speaker-specific Silence–Target–\linebreak Non-target–Overlap (STNO) masks to condition the ASR model, in regions with fully overlapped speech, these masks can become ambiguous, providing nearly identical conditioning even though the transcriptions for different speakers should differ. To address this, we introduce SE-DiCoW (Self-Enrolled Diarization-Conditioned Whisper), which resolves this ambiguity by automatically selecting the best available segments of the target speaker's speech based on diarization outputs and incorporating them as additional conditioning examples via cross-attention.
In addition, we enhance the original DiCoW framework with improved model initialization, refined training data segmentation, and data augmentations. 
Combined with self-enrollment, these advances yield a substantially stronger system: on the EMMA MT-ASR benchmark\footnote{\url{https://huggingface.co/spaces/BUT-FIT/EMMA_leaderboard}}, SE-DiCoW\footnote{\url{https://huggingface.co/BUT-FIT/SE_DiCoW}} reduces macro-averaged tcpWER by 52.4\,\% over the original DiCoW\footnote{\url{https://huggingface.co/BUT-FIT/DiCoW_v1}} with oracle diarization, and with real diarization attains state-of-the-art performance on AMI SDM~\cite{Mccowan2005_ami} and Libri2Mix~\cite{Cosentino2020LibriMixAO} while remaining comparable to domain-tuned systems on other datasets.

\begin{figure*}[t]
    \centering
    \includegraphics[width=2.0\columnwidth]{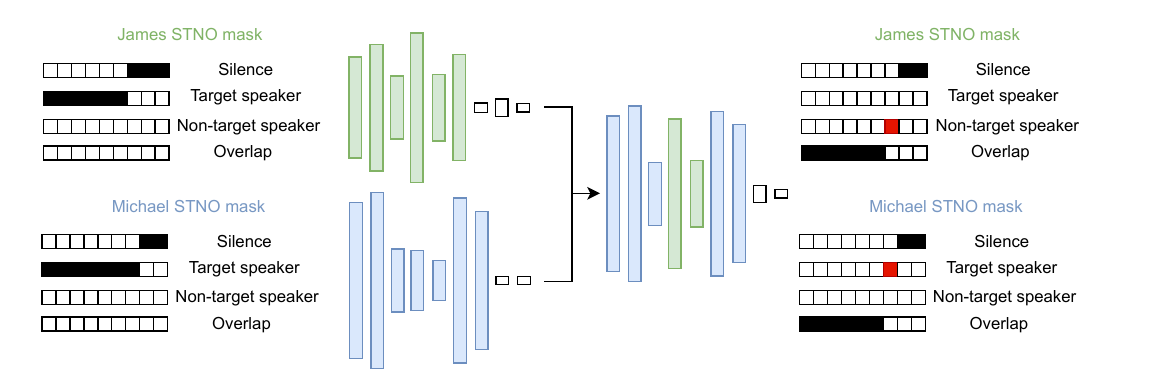}
    \caption{STNO ambiguity in highly overlapping speech regions. The STNO masks of James and Michael differ only at the positions highlighted in red, leaving a single (non-)target speaker frame for the model to exploit to track the target speaker.
    }
\label{fig:stno_problem}
\end{figure*}

\section{Method}
\label{sec:method}
This section reviews DiCoW and introduces our extensions. The enhanced SE-DiCoW architecture is shown in Figure~\ref{fig:dicow}, where the original DiCoW structure is also visible.

\subsection{DiCoW: Diarization-Conditioned Whisper}
DiCoW~\cite{DiCoW} builds upon the Whisper architecture to perform target-speaker ASR by conditioning directly on frame-by-frame speaker activity probabilities, $d(s, t)$, where $s$ indexes speakers and $t$ indexes time. This approach avoids explicit speaker identity modeling and enables generalization to unseen speakers. 

For a given target speaker $s_k$, DiCoW constructs a Silence–\linebreak Target–Non-target–Overlap ($\text{STNO}$) mask from the diarization output to capture four frame-level speech probabilities for: target speaker active, non-target active, overlap of target, or silence:
\begin{align}\label{eq:stno}
    p_{\mathcal{S}}^t  &= \prod_{s=1}^S (1 - d(s, t)), \quad
    p_{\mathcal{T}}^t  = d(s_k, t)  \prod_{\substack{s=1 \\ s \neq s_k}}^S (1 - d(s, t)) \nonumber \\
    p_{\mathcal{N}}^t  &= \left(1 - p_{\mathcal{S}}^t\right) - d(s_k, t), \quad
    p_{\mathcal{O}}^t  = d(s_k, t) - p_{\mathcal{T}}^t.
\end{align}

Instead of directly masking the input audio, DiCoW integrates STNO masks through Frame-Level Diarization-Dependent Transformations (FDDT), which modulate the internal representations of each Transformer layer. Each layer is augmented with four learnable affine transformation matrices\footnote{Throughout this work, we restrict these matrices to diagonal form.}, $(\mathbf{W}_i^l, \mathbf{b}_i^l)$, corresponding to the four STNO categories $i \in \{\mathcal{S}, \mathcal{T}, \mathcal{N}, \mathcal{O}\}$. The input to the Transformer encoder block at layer $l$ and frame $t$ is transformed as a probabilistic blend of corresponding transformations, weighted by the STNO probabilities:
\begin{equation}
\label{eq:FDDT_summary}
    \hat{\mathbf{z}}^l_t = \sum_{i \in \{\mathcal{S}, \mathcal{T}, \mathcal{N}, \mathcal{O}\}} (\mathbf{W}^l_i \mathbf{z}^l_t + \mathbf{b}^l_i)p^t_i.
\end{equation}

\subsection{Self-Enrolled Diarization-Conditioned Whisper}
\label{subsec:se-dicow}

Despite DiCoW's success, a critical limitation arises in fully overlapped speech regions, where different target speakers may receive nearly identical STNO conditioning. This makes it difficult for the model to distinguish speakers and produce accurate transcriptions (see Figure~\ref{fig:stno_problem}). Such ambiguity fundamentally limits the model's ability to maintain speaker-specific context in highly challenging scenarios, including recordings with multiple simultaneous conversations, as expected in the CHiME-9 MCoRec Challenge\footnote{\url{https://www.chimechallenge.org/current/task1/index}}.

To address this limitation, we introduce a \emph{self-enrollment mechanism} that automatically selects the most relevant reference segment of the target speaker within a recording. The model scans the entire recording $\mathbf{R}$ to identify a segment $[t_{\text{start}}, t_{\text{end}}]$ of fixed length\footnote{SE-DiCoW operates under Whisper’s long-form sequential decoding, processing the recording in 30\,s windows.} that maximizes the sum of target speaker probabilities $p_{\mathcal{T}}^t$, derived during inference from the diarization output $d(s,t)$:
\begin{equation}
    [t_{\text{start}}, t_{\text{end}}] = \arg\max_{t_{\text{start}}, t_{\text{end}}} \sum_{t=t_{\text{start}}}^{t_{\text{end}}} p_{\mathcal{T}}^t.
\end{equation}

This self-enrollment segment is then incorporated as additional conditioning via cross-attention at each encoder layer $l$. Let $\mathbf{Z}^{(l)} = [\mathbf{z}_1^l, \mathbf{z}_2^l, \dots, \mathbf{z}_T^l]$ denote the sequence of hidden representations at layer $l$. The cross-attention mechanism operates as follows:
\begin{align}
\mathbf{Z}_{\text{se}}^{(l)} &= \text{EncoderLayer}^{(l)}(\mathbf{Z}_{\text{se}}^{(l-1)}, \text{STNO}_{\text{se}}) \\
\mathbf{C}^{(l)} &= \text{CrossAttention}(\mathbf{Q} = \mathbf{Z}^{(l-1)}, \mathbf{K} = \mathbf{Z}_{\text{se}}^{(l)}, \mathbf{V} = \mathbf{Z}_{\text{se}}^{(l)}) \\
\mathbf{Z}_{\text{aug}}^{(l)} &= \text{MLP}([\mathbf{Z}^{(l-1)}; \mathbf{C}^{(l)}]) + \mathbf{Z}^{(l-1)} \\
\mathbf{Z}^{(l)} &= \text{EncoderLayer}^{(l)}(\mathbf{Z}_{\text{aug}}^{(l)}, \text{STNO}),
\end{align}
where $\mathbf{Q}, \mathbf{K}, \mathbf{V}$ stand for the query, key, and value matrices used in cross-attention, $[\mathbf{Z}^{(l-1)}; \mathbf{C}^{(l)}]$ denotes concatenation along the feature dimension, and the MLP is a 2-layer feedforward network. The processing of the input mixture $\mathbf{X}$, conditioned on its corresponding STNO mask and the self-enrollment segment $\mathbf{X}_{\text{se}}$ together with $\text{STNO}_{\text{se}}$, is illustrated in the Figure~\ref{fig:dicow}. Newly added modules are highlighted in red. Loss is computed only on the representations of $\mathbf{X}$ and not on those coming from the self-enrollment segment $\mathbf{X}_{\text{se}}$. This mechanism enables the model to maintain consistent speaker-specific representations even when the STNO masks are ambiguous, as illustrated in Figure~\ref{fig:stno_problem}.

\subsection{Additional DiCoW Improvements}\label{subsec:modifications}
Beyond the self-enrollment mechanism, we introduce several ad-hoc refinements to improve system performance. The model without self-enrollment, released as DiCoW v3.3\footnote{\url{https://huggingface.co/BUT-FIT/DiCoW_v3_3}}, represents an upgraded variant of the original DiCoW.

\textbf{Pre-Positional Embedding FDDT Layer:}
We introduce an additional FDDT module immediately after the convolutional subsampling and before summing with the positional embedding, as highlighted in Figure~\ref{fig:dicow}. In contrast, the original DiCoW applies the first FDDT only after the sequence has been augmented with positional embeddings. This new layer uses the same STNO conditioning mechanism as in Eq.~\eqref{eq:FDDT_summary}. To mitigate overly aggressive suppression, we increase the initialization diagonal scaling factor~\cite{DiCoW} from 0.1 to 0.5 for the non-target and silence transformation matrices. For further details on FDDT initialization, see Section 4.4 of our prior work~\cite{DiCoW}.

\textbf{Data Augmentations:} To improve robustness against diarization errors, we apply Gaussian noise $\epsilon^t \sim \mathcal{N}(0, 0.2^2)$ to STNO masks with probability 0.75, followed by re-normalization:
\[
\tilde{p}^t = \frac{\max\left(p^t + \epsilon^t, 0\right)}{\sum_{i} \max\left(p^t_i + \epsilon^t_i, 0\right)}
\]
Additional augmentations include segment-wise STNO most-likely-class activity flips, applied to each training sample with probability 0.3. The recording is divided into segments with lengths sampled uniformly from $[0.1, 1.0]$\,s, and for each segment, the dominant class is flipped independently with probability 0.1. Further, we apply SpecAugment jointly to the concatenated input signal and STNO mask~\cite{park_specaugment_2019}, and add MUSAN noises~\cite{musan2015} with probability 0.3.

\textbf{Corrected Training Data Segmentation:} In prior work~\cite{DiCoW}, 30\,s training segments were prepared so that each segment ended with an explicit end-of-segment timestamp, which differs from Whisper’s original training data. We corrected this by not enforcing an end timestamp when an utterance extends beyond the 30\,s window; such segment labels are now terminated solely with the end-of-sequence (EOS) token.

\section{Experimental Setup}
We follow the experimental protocol of the DiCoW paper, training on a mixture of AMI~\cite{Mccowan2005_ami}, NOTSOFAR-1\footnote{We use version 240825.1, a subset of the original challenge dataset.}~\cite{vinnikov24_interspeech}, and Libri2Mix/\linebreak 3Mix~\cite{Cosentino2020LibriMixAO}. In addition, we synthesize extra training mixtures from LibriSpeech~\cite{librispeech} by randomly overlapping up to three segments with partial overlap ratios sampled uniformly in the range $[0.8, 1.0]$. For the LibriSpeech-based data, we construct on-the-fly enrollment mixtures $\mathbf{X}_{\text{se}}$ by mixing three segments: one from the target speaker (not used in the input mixture) and two from other speakers. The target speaker segment is overlapped with the others with an overlap ratio sampled uniformly from $[0.3, 1.0]$, which can result in fully overlapped signals. This overlap distribution mimics conditions observed in real datasets such as AMI or NOTSOFAR-1.

All models use Whisper-large-v3-turbo\footnote{\url{https://huggingface.co/openai/whisper-large-v3-turbo}} as the backbone, fine-tuned with a learning rate of $2\times10^{-6}$, batch size of 96, 2k warm-up steps, and 40k total training steps under a cosine decay schedule. Performance is measured using time-constrained minimum-permutation WER (tcpWER)~\cite{Neumann2023MeetEval} with a 5\,s collar. Both the training\footnote{\url{https://github.com/BUTSpeechFIT/TS-ASR-Whisper}} and inference\footnote{\url{https://github.com/BUTSpeechFIT/DiCoW}} codes are publicly available.

We report results under two diarization conditions. First, oracle diarization, which uses reference speaker activity to construct STNO masks, provides an upper bound on achievable performance. Second, real diarization, where STNO masks are derived from DiariZen, a Pyannote-style diarization system~\cite{han25_interspeech}, which serves as a state-of-the-art diarization front-end.
For evaluation, we include multiple domains and recording conditions. On AMI, we report results on both the SDM (single distant microphone) and IHM-Mix (mixture of individual headset microphones) settings. On LibriSpeechMix~\cite{kanda20b_interspeech}, we evaluate mixtures of 1, 2, and 3 speakers. Unlike Libri2Mix/3Mix, these mixtures do not contain fully overlapped speech and better mimic real-world conversational patterns.

\begin{table*}[ht]
    \centering
\caption{tcpWER (\%) (5\,s collar) on real and synthetic datasets using oracle and DiariZen diarization. SE-DiCoW consistently yields the lowest error rates, especially in high-overlap conditions. {\color{darkgray}\emph{Dark grey}} indicates degradation caused by DiariZen's limit of two concurrent speakers. ($^*$) Denotes models trained on the original NOTSOFAR-1 dataset, which is a superset of the currently public release (containing the restricted Dev-set-2 and using Dev-set-1 for training).}
    \label{tab:emma_tcp}
    \setlength{\tabcolsep}{2pt} 
    \begin{tabular}{l*{10}c}
        \toprule
        & \multicolumn{1}{c}{NOTSOFAR-1} 
        & \multicolumn{2}{c}{AMI} 
        & \multicolumn{3}{c}{LibriSpeechMix} 
        & \multicolumn{2}{c}{Libri2Mix} 
        & \multicolumn{2}{c}{Libri3Mix} \\
        & Small-SDM & SDM & IHM-Mix & 1 & 2 & 3 & Both & Clean & Both & Clean \\
        \midrule
        & \multicolumn{10}{c}{oracle diarization}  \\
        \midrule     
        $\text{DiCoW}^{*}$ & 19.6 & 17.5 & 13.7 & 1.8 & 6.1 & 14.6 & 15.9 & 6.9 & 49.1 & 39.5 \\
        + flexible data seg. & 17.6 & 16.0 & 12.5 & 1.9 & 3.3 & 7.1 & 14.0 & 6.6 & 45.2 & 35.9 \\
        + new model init.. & 16.6 & 15.4 & 12.8 & 1.8 & 3.5 & 7.1 & 10.4 & 4.0 & 39.6 & 29.1 \\
        + augmentations [DiCoW v3.3] & 16.0 & 14.5 & \textbf{11.0} &\textbf{ 1.7} & \textbf{2.1} & 3.3 & 9.7 & 3.8 & 27.7 & 16.0  \\
        SE-DiCoW & \textbf{15.8} & \textbf{14.3} & \textbf{11.0} & \textbf{1.7} & \textbf{2.1} & \textbf{2.9} & \textbf{7.7} & \textbf{2.8} & \textbf{19.9} & \textbf{9.7} \\
        \midrule
        & \multicolumn{10}{c}{DiariZen diarization}  \\
        DER [0.25\,s collar]       & 12.6 & 10.4 & 21.9 & 12.2 & 13.2 & {\color{darkgray} 21.7} & 9.1 & 12.9 & {\color{darkgray} 27.6} & {\color{darkgray} 29.6}  \\
        MSCE & 0.37 & 0.38 & 0.69 & 0.0 & 0.01 & {\color{darkgray} 0.67} & 0.0 & 0.0 & {\color{darkgray} 0.99} & {\color{darkgray} 0.94} \\
        \midrule
        $\text{DiCoW}^{*}$      & 29.8 & 21.4 & 17.0 & 6.4 & 17.9 & {\color{darkgray} 32.5} & 21.6 & 16.1 & {\color{darkgray} 52.4} & {\color{darkgray} 47.1} \\
       DiCoW v3.3 & 26.6 & 18.6 & 15.2 & \textbf{1.8} & 3.1 & {\color{darkgray}21.7} & 9.7 & 3.5 & {\color{darkgray}38.6} & {\color{darkgray}31.6}\\
        SE-DiCoW   & 26.1 & \textbf{18.5} & 15.3 & \textbf{1.8} & 3.0 & {\color{darkgray}21.1} & \textbf{8.7} & \textbf{3.4} & {\color{darkgray}35.6} & {\color{darkgray}29.3} \\
       SOTA (as of September 2025) & $\mathbf{23.6}^{*}$~\cite{niu24_chime} & 21.2~\cite{kanda21_interspeech} & \textbf{14.9}~\cite{kanda21_interspeech} & 2.2~\cite{wang25y_interspeech} & \textbf{2.8}~\cite{wang25y_interspeech} & \textbf{5.0}~\cite{wang25y_interspeech} & 9.2~\cite{shi2025serializedoutputpromptinglarge} & 3.6~\cite{shi2025serializedoutputpromptinglarge} & \textbf{28.1}~\cite{shi2025serializedoutputpromptinglarge} & \textbf{16.5}~\cite{shi2025serializedoutputpromptinglarge} \\ 
        \bottomrule        
    \end{tabular}
\end{table*}

\section{Results}
Table~\ref{tab:emma_tcp} reports tcpWER across both real and synthetic multi-speaker benchmarks. All results are obtained using the CHiME-8 text normalization and follow the evaluation protocol of the EMMA MT-ASR Benchmark. In addition to ASR performance, we also report the diarization error rate (DER)\footnote{Reported DER is computed using segment-level annotations, with a 0.25\,s collar applied to mitigate errors introduced by imprecise labels.} of DiariZen and the corresponding mean speaker counting error (MSCE), computed as\linebreak
$
\text{MSCE} = \frac{1}{N} \sum_{r=1}^{N} \lvert C_r - C_h \rvert,
$
where $N$ is the number of recordings in the dataset, $C_r$ is the reference number of speakers, and $C_h$ is the number of speakers inferred by the diarization system.

We first evaluate oracle diarization to set upper bounds. While competitive, baseline DiCoW falters in heavy overlap—most notably in Libri3Mix-both, where three LibriSpeech recordings are mixed without temporal offsets, creating a scenario challenging even for humans.

Corrected training data segmentation yields consistent improvements, particularly on AMI and NOTSOFAR-1, where long-form sequential decoding is employed. Refinements to model initialization further reduce error rates, and data augmentation provides additional gains. Consequently, DiCoW v3.3 shows further reductions in tcpWER across all evaluated benchmarks.

SE-DiCoW outperforms all other variants, achieving the lowest tcpWER across all benchmarks. On Libri3Mix-clean, SE-DiCoW reduces error by more than 75\% relative to the original DiCoW. Crucially, these improvements are not limited to fully overlapped synthetic data; absolute tcpWER reductions of 0.2 are also observed on NOTSOFAR-1 and AMI-SDM. These results demonstrate that self-enrollment effectively resolves STNO ambiguity in overlapped regions and, combined with improved initialization, segmentation, and augmentation, produces a state-of-the-art system for TS-ASR.

When moving from oracle to real diarization with DiariZen, performance degrades noticeably across datasets. Nevertheless, SE-DiCoW remains comparable to state-of-the-art approaches, each fine-tuned for a specific domain and typically evaluated using WER or cpWER—both of which represent lower bounds on tcpWER. The degradation is particularly pronounced on datasets with more than two simultaneously overlapping speakers, reflecting a limitation of the DiariZen~\cite{han25_interspeech} system, which models a powerset of 11 classes with at most two active speakers at a time~\cite{plaquet23_interspeech}. This issue is most evident in Libri3Mix, where the mean speaker counting error indicates that one speaker is consistently missing.

\subsection{Analysis of Self-Enrollment Mixture Composition}
\begin{table}[t]
    \centering
    \setlength{\tabcolsep}{4pt} 
    \caption{Analysis of self-enrollment mixture composition on Libri3Mix Clean test set. tcpWER (\%) is reported for different numbers of speakers in the enrollment segment and varying overlap ratios with the target speaker.}
    \label{tab:enrollment_analysis}
    \begin{tabular}{lccccc}
        \toprule
        \multirow{2}{*}{Enrollment Composition} & \multicolumn{5}{c}{Overlap Ratio w/ Target Speaker} \\
        & 0\% & 25\% & 50\% & 75\% & 100\% \\
        \midrule
        Target speaker only & 9.67   & -- & -- & -- & -- \\
        Target + 1 interferer & -- & 9.66 & 9.68 & 9.66 & 9.67 \\
        Target + 2 interferers & -- & \textbf{9.61} & 9.72 & 9.73 & 9.87 \\
        \bottomrule
    \end{tabular}
    
\end{table}
\vspace{-0.2cm}
In Table~\ref{tab:emma_tcp}, SE-DiCoW was evaluated with enrollment overlap ratios sampled from $\mathcal{U}[0.3, 1.0]$, reflecting real conversational conditions. To analyze the effect of enrollment composition, we performed a controlled study on Libri3Mix clean by varying (1) the number of concurrent speakers and (2) the overlap ratio with the target speaker.

Table~\ref{tab:enrollment_analysis} shows SE-DiCoW works best with 3 speakers (Target + 2 interferers) and minimal overlap (25\%), achieving the lowest error rate of \textbf{9.61\%}. This appears to be the best scenario because the model can utilize context to learn what the target speaker sounds like when slightly overlapped. Notable degradation is observed only when the segment is fully overlapped with too many speakers: while performance remains stable at 9.87\% with 3 speakers, it degrades to 12.2\% and 12.4\% with 4 and 5 speakers, respectively. Nevertheless, SE-DiCoW still significantly outperforms the baseline DiCoW.
These results highlight SE-DiCoW’s practicality: even when clean segments are unavailable, the self-enrollment mechanism naturally selects regions with a high proportion of frames having large $p_{\mathcal{T}}^t$ values, thereby favoring cleaner references while preserving robustness in more challenging cases.

\section{Conclusion}
We introduced SE-DiCoW (Self-Enrolled Diarization-Conditioned Whisper), which addresses a key limitation of the original DiCoW: ambiguity in STNO conditioning during fully overlapped speech. By automatically selecting target-speaker reference segments and incorporating them via cross-attention, SE-DiCoW effectively resolves speaker disambiguation when different speakers receive nearly identical conditioning.  
Comprehensive evaluation shows substantial gains across the diverse datasets of the EMMA MT-ASR benchmark. SE-DiCoW reduces macro-average tcpWER by 52.4\% over DiCoW, with over 75\% relative improvement on Libri3Mix clean and consistent gains on real conversational data. Enrollment analysis further demonstrates robustness to imperfect reference enrollment segments, underscoring its practicality in real-world settings.  
In addition to self-enrollment, improvements in initialization, data segmentation, and augmentation contribute to overall effectiveness.  
The resulting framework achieves performance on par with the best domain-tuned systems reported in the literature, while preserving DiCoW’s strong cross-domain generalization.  
Future work will focus on jointly fine-tuning diarization and TS-ASR within a unified framework, aiming to mitigate the degradation observed with inferred diarization (Table~\ref{tab:emma_tcp}).

\newpage

\section{Acknowledgements}
This work, done at JSALT 2025, was partially supported by Ministry of Education, Youth and Sports of the Czech Republic (MoE) through the OP JAK project ``Linguistics, Artificial Intelligence and Language and Speech Technologies: from Research to Applications'' (ID:CZ.02.01.01/00/23\_020/0008518), Brno Ph.D. Talent Scholarship Programme, and by Johns Hopkins University via corporate gifts. Computing on IT4I supercomputer was supported by MoE through the e-INFRA CZ (ID:90254).
\vspace{-5pt}
\bibliographystyle{IEEEbib}
\bibliography{refs}

@article{DiCoW,
title = {{DiCoW}: Diarization-conditioned {Whisper} for target speaker automatic speech recognition},
journal = {Computer Speech \& Language},
volume = {95},
pages = {101841},
year = {2026},
issn = {0885-2308},
doi = {https://doi.org/10.1016/j.csl.2025.101841},
url = {https://www.sciencedirect.com/science/article/pii/S088523082500066X},
author = {Alexander Polok and Dominik Klement and Martin Kocour and Jiangyu Han and Federico Landini and Bolaji Yusuf and Matthew Wiesner and Sanjeev Khudanpur and Jan Černocký and Lukáš Burget},
}

@inproceedings{han25_interspeech,
  title     = {Fine-tune Before Structured Pruning: Towards Compact and Accurate Self-Supervised Models for Speaker Diarization},
  author    = {Jiangyu Han and Federico Landini and Johan Rohdin and Anna Silnova and Mireia Diez and Jan Černocký and Lukáš Burget},
  year      = {2025},
  booktitle = {{Interspeech 2025}},
  pages     = {1583--1587},
  doi       = {10.21437/Interspeech.2025-484},
  issn      = {2958-1796},
}

@inproceedings{cornell2023chime,
  title     = {The {CHiME-7} {DASR} Challenge: Distant Meeting Transcription with Multiple Devices in Diverse Scenarios},
  author    = {Samuele Cornell and Matthew S. Wiesner and Shinji Watanabe and Desh Raj and Xuankai Chang and Paola Garcia and Yoshiki Masuyam and Zhong-Qiu Wang and Stefano Squartini and Sanjeev Khudanpur},
  year      = {2023},
  booktitle = {7th International Workshop on Speech Processing in Everyday Environments (CHiME 2023)},
  pages     = {1--6},
  doi       = {10.21437/CHiME.2023-1},
}

@inproceedings{cornell2024one,
  title={One Model to Rule Them All? {Towards} End-to-End Joint Speaker Diarization and Speech Recognition},
  author={Cornell, Samuele and Jung, Jee-weon and Watanabe, Shinji and Squartini, Stefano},
  booktitle={2024 IEEE International Conference on Acoustics, Speech and Signal Processing (ICASSP)},
  pages={11856--11860},
  year={2024},
  organization={IEEE}
}

@inproceedings{cornell2024chime,
  title={The {CHiME-8} {DASR} Challenge for Generalizable and Array Agnostic Distant Automatic Speech Recognition and Diarization},
  author    = {Samuele Cornell and Tae Jin Park and He Huang and Christoph Boeddeker and Xuankai Chang and Matthew Maciejewski and Matthew S Wiesner and Paola Garcia and Shinji Watanabe},
  year      = {2024},
  booktitle = {8th International Workshop on Speech Processing in Everyday Environments (CHiME 2024)},
  pages     = {1--6},
  doi       = {10.21437/CHiME.2024-1},
}

@article{Cosentino2020LibriMixAO,
  title={{LibriMix}: An Open-Source Dataset for Generalizable Speech Separation},
  author={Joris Cosentino and Manuel Pariente and Samuele Cornell and Antoine Deleforge and Emmanuel Vincent},
  journal={arXiv: Audio and Speech Processing},
  year={2020},
  url={https://api.semanticscholar.org/CorpusID:218862876}
}

@inproceedings{Kanda2019_spkloss,
  title     = {Auxiliary Interference Speaker Loss for Target-Speaker Speech Recognition},
  author    = {Naoyuki Kanda and Shota Horiguchi and Ryoichi Takashima and Yusuke Fujita and Kenji Nagamatsu and Shinji Watanabe},
  year      = {2019},
  booktitle = {Interspeech 2019},
  pages     = {236--240},
  doi       = {10.21437/Interspeech.2019-1126},
  issn      = {2958-1796},
}

@inproceedings{kanda20b_interspeech,
  title     = {Serialized Output Training for End-to-End Overlapped Speech Recognition},
  author    = {Naoyuki Kanda and Yashesh Gaur and Xiaofei Wang and Zhong Meng and Takuya Yoshioka},
  year      = {2020},
  booktitle = {Interspeech 2020},
  pages     = {2797--2801},
  doi       = {10.21437/Interspeech.2020-999},
  issn      = {2958-1796},
}

@inproceedings{ma2024extending,
  title={Extending {Whisper} with prompt tuning to target-speaker {ASR}},
  author={Ma, Hao and Peng, Zhiyuan and Shao, Mingjie and Li, Jing and Liu, Ju},
  booktitle={2024 IEEE International Conference on Acoustics, Speech and Signal Processing (ICASSP)},
  pages={12516--12520},
  year={2024},
  organization={IEEE}
}

@article{Mccowan2005_ami,
author = {Mccowan, Iain and Carletta, J and Kraaij, Wessel and Ashby, Simone and Bourban, S and Flynn, M and Guillemot, M and Hain, Thomas and Kadlec, J and Karaiskos, V and Kronenthal, M and Lathoud, Guillaume and Lincoln, Mike and Lisowska Masson, Agnes and Post, Wilfried and Reidsma, Dennis and Wellner, P},
year = {2005},
month = {01},
pages = {},
title = {The {AMI} meeting corpus},
journal = {Int'l. Conf. on Methods and Techniques in Behavioral Research}
}

@inproceedings{Neumann2023MeetEval,
  author    = {T. v. Neumann and C. B. Boeddeker and M. Delcroix and R. Haeb-Umbach},
  title     = {{MeetEval}: A Toolkit for Computation of Word Error Rates for Meeting Transcription Systems},
  booktitle = {Proceedings of the 7th International Workshop on Speech Processing in Everyday Environments (CHiME 2023)},
  pages     = {27--32},
  year      = {2023},
  doi       = {10.21437/CHiME.2023-6},
}

@inproceedings{niu24_chime,
  title     = {The {USTC-NERCSLIP} Systems for the {CHiME-8 NOTSOFAR-1} Challenge},
  author    = {Shutong Niu and Ruoyu Wang and Jun Du and Gaobin Yang and Yanhui Tu and Siyuan Wu and Shuangqing Qian and Huaxin Wu and Haitao Xu and Xueyang Zhang and Guolong Zhong and Xindi Yu and Jieru Chen and Mengzhi Wang and Di Cai and Tian Gao and Genshun Wan and Feng Ma and Jia Pan and Jianqing Gao},
  year      = {2024},
  booktitle = {8th International Workshop on Speech Processing in Everyday Environments (CHiME 2024)},
  pages     = {31--36},
  doi       = {10.21437/CHiME.2024-7},
}

@INPROCEEDINGS{peng2023owsm,
  author={Peng, Yifan and Tian, Jinchuan and Yan, Brian and Berrebbi, Dan and Chang, Xuankai and Li, Xinjian and Shi, Jiatong and Arora, Siddhant and Chen, William and Sharma, Roshan and Zhang, Wangyou and Sudo, Yui and Shakeel, Muhammad and Jung, Jee-Weon and Maiti, Soumi and Watanabe, Shinji},
  booktitle={2023 IEEE Automatic Speech Recognition and Understanding Workshop (ASRU)},
  title={Reproducing {Whisper}-Style Training Using An Open-Source Toolkit And Publicly Available Data},
  year={2023},
  volume={},
  number={},
  pages={1-8},
  doi={10.1109/ASRU57964.2023.10389676}
}

@inproceedings{plaquet23_interspeech,
  title     = {Powerset multi-class cross entropy loss for neural speaker diarization},
  author    = {Alexis Plaquet and Hervé Bredin},
  year      = {2023},
  booktitle = {Interspeech 2023},
  pages     = {3222--3226},
  doi       = {10.21437/Interspeech.2023-205},
  issn      = {2958-1796},
}

@inproceedings{polok24_butjhu,
  title     = {{BUT/JHU} System Description for {CHiME-8 NOTSOFAR-1} Challenge},
  author    = {Alexander Polok and Dominik Klement and Jiangyu Han and Simon Sedl\'a\v{c}ek and Bolaji Yusuf and Matthew Maciejewski and Matthew Wiesner and Luk\'a\v{s} Burget},
  year      = {2024},
  booktitle = {8th International Workshop on Speech Processing in Everyday Environments (CHiME 2024)},
  pages     = {18--22},
  doi       = {10.21437/CHiME.2024-4},
}

@INPROCEEDINGS{polok2024targetspeakerasrwhisper,
  author={Polok, Alexander and Klement, Dominik and Wiesner, Matthew and Khudanpur, Sanjeev and Černocký, Jan and Burget, Lukáš},
  booktitle={2025 IEEE International Conference on Acoustics, Speech and Signal Processing (ICASSP)},
  title={Target Speaker {ASR} with {Whisper}},
  year={2025},
  volume={},
  number={},
  pages={1-5},
  doi={10.1109/ICASSP49660.2025.10887683}}

@inproceedings{radford2023robust,
  title={Robust speech recognition via large-scale weak supervision},
  author={Radford, Alec and Kim, Jong Wook and Xu, Tao and Brockman, Greg and McLeavey, Christine and Sutskever, Ilya},
  booktitle={International conference on machine learning},
  pages={28492--28518},
  year={2023},
  organization={PMLR}
}

@inproceedings{vinnikov24_interspeech,
  title     = {{NOTSOFAR-1} Challenge: New Datasets, Baseline, and Tasks for Distant Meeting Transcription},
  author    = {Alon Vinnikov and Amir Ivry and Aviv Hurvitz and Igor Abramovski and Sharon Koubi and Ilya Gurvich and Shai Peer and Xiong Xiao and Benjamin Martinez Elizalde and Naoyuki Kanda and Xiaofei Wang and Shalev Shaer and Stav Yagev and Yossi Asher and Sunit Sivasankaran and Yifan Gong and Min Tang and Huaming Wang and Eyal Krupka},
  year      = {2024},
  booktitle = {Interspeech 2024},
  pages     = {5003--5007},
  doi       = {10.21437/Interspeech.2024-1788},
  issn      = {2958-1796},
}

@INPROCEEDINGS{Zili23_adapting,
  author={Huang, Zili and Raj, Desh and García, Paola and Khudanpur, Sanjeev},
  booktitle={2023 IEEE International Conference on Acoustics, Speech and Signal Processing (ICASSP)},
  title={Adapting Self-Supervised Models to Multi-Talker Speech Recognition Using Speaker Embeddings},
  year={2023},
  volume={},
  number={},
  pages={1-5},
  doi={10.1109/ICASSP49357.2023.10097139}
}

@INPROCEEDINGS{librispeech,
  author={Panayotov, Vassil and Chen, Guoguo and Povey, Daniel and Khudanpur, Sanjeev},
  booktitle={2015 IEEE International Conference on Acoustics, Speech and Signal Processing (ICASSP)},
  title={Librispeech: An {ASR} corpus based on public domain audio books},
  year={2015},
  volume={},
  number={},
  pages={5206-5210},
  doi={10.1109/ICASSP.2015.7178964}
}

@inproceedings{puvvada24_interspeech,
  title     = {Less is More: Accurate Speech Recognition \& Translation without Web-Scale Data},
  author    = {Krishna C. Puvvada and Piotr Żelasko and He Huang and Oleksii Hrinchuk and Nithin Rao Koluguri and Kunal Dhawan and Somshubra Majumdar and Elena Rastorgueva and Zhehuai Chen and Vitaly Lavrukhin and Jagadeesh Balam and Boris Ginsburg},
  year      = {2024},
  booktitle = {{Interspeech 2024}},
  pages     = {3964--3968},
  doi       = {10.21437/Interspeech.2024-2294},
  issn      = {2958-1796},
}

@misc{polok2025mlcslmchallenge,
      title={{BUT} System for the {MLC-SLM} Challenge}, 
      author={Alexander Polok and Jiangyu Han and Dominik Klement and Samuele Cornell and Jan Černocký and Lukáš Burget},
      year={2025},
      eprint={2506.13414},
      archivePrefix={arXiv},
      primaryClass={eess.AS},
      url={https://arxiv.org/abs/2506.13414}, 
}

@inproceedings{wang25y_interspeech,
  title     = {Speaker Targeting via Self-Speaker Adaptation for Multi-talker {ASR}},
  author    = {Weiqing Wang and Taejin Park and Ivan Medennikov and Jinhan Wang and Kunal Dhawan and He Huang and Nithin {Rao Koluguri} and Jagadeesh Balam and Boris Ginsburg},
  year      = {2025},
  booktitle = {{Interspeech 2025}},
  pages     = {5498--5502},
  doi       = {10.21437/Interspeech.2025-2142},
  issn      = {2958-1796},
}

@INPROCEEDINGS{SALM,
  author={Chen, Zhehuai and Huang, He and Andrusenko, Andrei and Hrinchuk, Oleksii and Puvvada, Krishna C. and Li, Jason and Ghosh, Subhankar and Balam, Jagadeesh and Ginsburg, Boris},
  booktitle={ICASSP 2024 - 2024 IEEE International Conference on Acoustics, Speech and Signal Processing (ICASSP)}, 
  title={{SALM}: Speech-Augmented Language Model with in-Context Learning for Speech Recognition and Translation}, 
  year={2024},
  volume={},
  number={},
  pages={13521-13525},
  doi={10.1109/ICASSP48485.2024.10447553}}

@misc{musan2015,
  author = {David Snyder and Guoguo Chen and Daniel Povey},
  title = {{MUSAN}: A Music, Speech, and Noise Corpus},
  year = {2015},
  eprint = {1510.08484},
  note = {arXiv:1510.08484v1}
}

@inproceedings{park_specaugment_2019,
    title = {{SpecAugment}: A Simple Data Augmentation Method for Automatic Speech Recognition},
    shorttitle = {{SpecAugment}},
    url = {https://www.isca-archive.org/interspeech_2019/park19e_interspeech.html},
    doi = {10.21437/Interspeech.2019-2680},
    language = {en},
    urldate = {2024-02-23},
    booktitle = {Interspeech 2019},
    publisher = {ISCA},
    author = {Park, Daniel S. and Chan, William and Zhang, Yu and Chiu, Chung-Cheng and Zoph, Barret and Cubuk, Ekin D. and Le, Quoc V.},
    month = sep,
    year = {2019},
    pages = {2613--2617},
}

@inproceedings{yu17b_interspeech,
  title     = {Recognizing Multi-Talker Speech with Permutation Invariant Training},
  author    = {Dong Yu and Xuankai Chang and Yanmin Qian},
  year      = {2017},
  booktitle = {Interspeech 2017},
  pages     = {2456--2460},
  doi       = {10.21437/Interspeech.2017-305},
  issn      = {2958-1796},
}

@misc{shi2025serializedoutputpromptinglarge,
      title={Serialized Output Prompting for Large Language Model-based Multi-Talker Speech Recognition}, 
      author={Hao Shi and Yusuke Fujita and Tomoya Mizumoto and Lianbo Liu and Atsushi Kojima and Yui Sudo},
      year={2025},
      eprint={2509.04488},
      archivePrefix={arXiv},
      primaryClass={cs.CL},
      url={https://arxiv.org/abs/2509.04488}, 
}

@inproceedings{kanda21_interspeech,
  title     = {Large-Scale Pre-Training of End-to-End Multi-Talker {ASR} for Meeting Transcription with Single Distant Microphone},
  author    = {Naoyuki Kanda and Guoli Ye and Yu Wu and Yashesh Gaur and Xiaofei Wang and Zhong Meng and Zhuo Chen and Takuya Yoshioka},
  year      = {2021},
  booktitle = {Interspeech 2021},
  pages     = {3430--3434},
  doi       = {10.21437/Interspeech.2021-102},
  issn      = {2958-1796},
}

@misc{kocour2025adaptingdiarizationconditionedwhisperendtoend,
      title={Adapting Diarization-Conditioned {Whisper} for End-to-End Multi-Talker Speech Recognition}, 
      author={Martin Kocour and Martin Karafiat and Alexander Polok and Dominik Klement and Lukáš Burget and Jan Černocký},
      year={2025},
      eprint={2510.03723},
      archivePrefix={arXiv},
      primaryClass={eess.AS},
      url={https://arxiv.org/abs/2510.03723}, 
}

@misc{he2025scalingmultitalkerasrspeakeragnostic,
      title={Scaling Multi-Talker {ASR} with Speaker-Agnostic Activity Streams}, 
      author={Xiluo He and Alexander Polok and Jesús Villalba and Thomas Thebaud and Matthew Maciejewski},
      year={2025},
      eprint={2510.03630},
      archivePrefix={arXiv},
      primaryClass={eess.AS},
      url={https://arxiv.org/abs/2510.03630}, 
}

\end{document}